**Optical manipulation of Berry phase in a solid-state spin qubit**

Christopher G. Yale*,1,2, F. Joseph Heremans*,1, Brian B. Zhou*,1, Adrian Auer[3], Guido Burkard[3], and David D. Awschalom†,1,2

1. Institute for Molecular Engineering, University of Chicago, Chicago, Illinois 60637, USA
2. Department of Physics, University of California, Santa Barbara, California 93106, USA
3. Department of Physics, University of Konstanz, D-78457 Konstanz, Germany

*These authors contributed equally to this work.

†Corresponding Author: awsch@uchicago.edu

**The phase relation between quantum states represents an essential resource for the storage and processing of quantum information. While quantum phases are commonly controlled dynamically by tuning energetic interactions, utilizing geometric phases that accumulate during cyclic evolution may offer superior robustness to noise. To date, demonstrations of geometric phase control in solid-state systems rely on microwave fields that have limited spatial resolution. Here, we demonstrate an all-optical method based on stimulated Raman adiabatic passage to accumulate a geometric phase, the Berry phase, in an individual nitrogen-vacancy (NV) center in diamond. Using diffraction-limited laser light, we guide the NV center's spin along loops on the Bloch sphere to enclose arbitrary Berry phase and characterize these trajectories through time-resolved state tomography. We investigate the limits of this control due to loss of adiabiaticity and decoherence, as well as its robustness to noise intentionally introduced into the experimental control parameters, finding its resilience to be independent of the amount of Berry phase enclosed. These techniques set the foundation for optical geometric manipulation in future implementations of photonic networks of solid state qubits linked and controlled by light.**

When a quantum mechanical system evolves slowly along a closed loop in its parameter space, a given eigenstate may acquire a phase consisting of both a dynamic and geometric contribution. First proposed by S. Pancharatnam[1] in his study of cyclic rotations of the polarization of light, and later generalized by M. V. Berry[2], this adiabatic geometric phase is determined solely by the geometry of the traversed loop, in contrast to the dynamic phase that accumulates from the energetics and travel time of the intervening state evolution. Since the Berry phase is proportional to the area enclosed by the path in parameter space, it is intrinsically resilient to noise that causes deviations to the path but conserve the total enclosed area[3,4]. Geometric control thus represents a promising avenue for constructing fault-tolerant quantum logic gates[5,6].

Control over geometric phases, occurring both when the cyclic evolution is traversed adiabatically[2] and non-adiabatically[7], has been demonstrated in a variety of physical platforms, including liquid nuclear magnetic resonance[8], trapped atoms[9], and more recently in the solid-state in superconducting qubits[10,11] and defect spins[12–14]. However, current implementations of geometric phase control in solid-state systems have utilized microwaves that are difficult to localize and thus concede the ability to selectively address nearby qubits without crosstalk. Here, we manipulate the Berry phase in a solid-state qubit using diffraction-limited resonant laser fields. This opens the possibility for independent manipulation of single qubits in photonic networks[15] and spin arrays[16] through geometric principles where fine control over the energetics is inessential. While a similar optical protocol has been recently realized using trapped calcium ions[17], our realization in the solid state offers potential integration into photonic platforms and harnesses larger energy scales to enable significantly faster adiabatic control.

The method we employ is based on proposals to accumulate geometric phases via stimulated Raman adiabatic passage (STIRAP)[18,19]. In STIRAP, two laser fields couple two levels $|\alpha\rangle$ and $|\beta\rangle$ to a single excited level $|\varepsilon\rangle$ in a lambda ($\Lambda$) configuration (Fig. 1a). The amplitude and phase relation between the two laser fields define a new zero-energy eigenstate in the interaction picture, known as the dark state $|D\rangle$, which is a superposition of the $|\alpha\rangle$ and $|\beta\rangle$ states that does not couple to the light fields due to destructive interference. By adiabatically adjusting the amplitude and phase of the optical fields, any initial dark state can be connected to any other final dark state, thus transporting the state arbitrarily on the Bloch sphere spanned by $|\alpha\rangle$ and $|\beta\rangle$. Traditionally, STIRAP has been utilized in atomic[20,21] and solid-state systems[22,23] for highly efficient population transfer between the states $|\alpha\rangle$ and $|\beta\rangle$ bypassing the potentially lossy excited state $|\varepsilon\rangle$ (pole to pole evolution, shown by the red gradient curve in Fig 1b). However, when STIRAP is extended to a full loop (red and blue gradient curves together in Fig 1b), the dark state returns to itself and accumulates a Berry phase, $\gamma_B$,[17–19] proportional to the solid angle enclosed on the Bloch sphere.

In this work, we demonstrate Berry phase control through the adiabatic passage of a dark state within a single negatively charged nitrogen-vacancy (NV) center in diamond, a defect consisting of a substitutional nitrogen adjacent to a vacant site in the diamond lattice. Due to its long-lived spin coherence extending to room temperature, this defect is a promising candidate for quantum information and nanoscale sensing[24]. The NV center level structure has a ground-state spin triplet and an excited-state orbital doublet, spin triplet, where transitions between these levels become sharp and optically addressable[25] at cryogenic temperatures (<20 K). As a result, resonant optical excitation of the nitrogen-vacancy center has been used to demonstrate a number of quantum optics protocols[26], including spin-photon entanglement[27,28] and coherence[29], all-optical control[23,30,31], photonically entangled spins[32], and teleportation of its spin state[33].

**Understanding STIRAP in the NV Center**

We exploit a natural $\Lambda$ system within the NV center level structure, at cryogenic temperatures ($T$ = 8 K), formed by its ground state m$_S$ = -1 and +1 spin states, or $|-1_g\rangle$ and $|+1_g\rangle$, coupled to the $|A_2\rangle$ spin-orbit excited state[23,27,28] (Fig. 1a). We tune this $\Lambda$ system into a non-degenerate configuration using a ~117 G external magnetic field along the NV center axis to Zeeman split $|-1_g\rangle$ and $|+1_g\rangle$ by 655 MHz. On-chip microwave control[24] enables rotations between the third ground state $|0_g\rangle$ and either $|-1_g\rangle$ or $|+1_g\rangle$. These microwaves are used only for the preparation and tomographic projection of the spin state (described in Refs. 30, 31, and Supplemental Section 2.2.2), but are not involved in the accumulation of the geometric phase. Instead, we address our Λ system using a narrow-line tunable 637 nm diode laser (470 THz) fiber coupled into an electro-optic modulator (EOM). Driving the EOM with a signal generator places frequency harmonics on the laser equivalent to the $|-1_g\rangle$ / $|+1_g\rangle$ splitting (655 MHz) (Fig. 1c). A phase quadrature modulator controlled by an arbitrary waveform generator governs the relative amplitude and phase relations among these harmonic sidebands on nanosecond timescales. From this, we achieve full Bloch sphere control over the resultant dark state, $|D\rangle$[30].

In our experiment, the red-shifted first harmonic of the laser is tuned to the $|+1_g\rangle$ to $|A_2\rangle$ transition with Rabi coupling strength $\Omega_{+1}(t)$, and the zeroth harmonic is tuned to the $|-1_g\rangle$ to $|A_2\rangle$ transition with Rabi coupling strength $\Omega_{-1}(t)$ (Fig. 1c). The two laser fields are deliberately detuned from the one-photon resonance by a red shift $\Delta \approx 65 \pm 15$ MHz to limit unintended absorption during STIRAP, while detuning from the two-photon resonance by $|\delta| < 150$ kHz arises due to experimental uncertainty (see dynamic phase discussion below). In our protocol, we begin by preparing the spin into $|-1_g\rangle$. We then apply only $\Omega_{+1}$ which sets $|-1_g\rangle$ as the dark state. By adiabatically shifting the relative intensity from $\Omega_{+1}$ to $\Omega_{-1}$, the dark state gradually moves on the surface of the Bloch sphere from $|-1_g\rangle$ to the opposite $|+1_g\rangle$ pole (Fig. 1d), with minimal absorption through the $|A_2\rangle$ excited state. The precise time evolution of the field amplitudes $\Omega_{+1}(t)$ and $\Omega_{-1}(t)$, displayed in Fig. 1d, is governed by the pulse shape applied to the EOM and the harmonic generation relation in an EOM (Supplemental Section 2.2.1). A phase shift, $\Phi$, between the optical fields at the opposite pole ($t = \frac{\tau}{2}$) and a reversal of the intensity shift returns the dark state to the $|-1_g\rangle$ pole along a different longitude on the Bloch sphere completing a cyclic route in traversal time $\tau$. This 'tangerine slice' trajectory with wedge angle, $\Phi$, circumscribes a solid angle, $2\Phi$, and gives rise to an accumulated Berry phase. (Fig. 1b,d)

Prior to investigating the Berry phase, we explore the mechanisms limiting STIRAP in the NV center by tomographically reconstructing the path of the spin on the $|-1_g\rangle$ to $|+1_g\rangle$ Bloch sphere (Fig. 2a). In this particular instance, we demonstrate a trajectory with outbound (red) and inbound (blue) paths separated by $\Phi$ = 120° for an adiabatic cycle time of $\tau$ = 1200 ns and a peak optical Rabi frequency $\Omega_R$ = $31 \pm 3$ MHz for the $\Omega_{-1}$ transition. From this time-resolved reconstruction, we observe that the length of the dark state Bloch vector (Fig. 2b) decreases around the equator, revives near the opposite pole, and ultimately returns to the initial pole with 65% of its original magnitude.

This decrease in the state magnitude along the traversal is due to a lag in the adiabatic following of the dark state. Increases in the velocity of the trajectory can cause coupling strengths between the dark and non-dark eigenstates to exceed their energy gap, leading to non-adiabatic evolution. These energy gaps decrease with weaker driving fields, $\Omega_R$. Non-adiabatic effects raise the likelihood of occupying non-dark states that include components of the excited state $|A_2\rangle$[20]. Occupation of these states reduces the overall magnitude of the state vector, as these states point in different directions with respect to the intended dark state in the $|-1_g\rangle$ / $|+1_g\rangle$ Bloch sphere. Furthermore, as they contain components in $|A_2\rangle$, absorption can occur and lead to decay either into the intended dark state, causing recovery of the magnitude, or into $|0_g\rangle$, causing irreversible loss out of the subpace. For our STIRAP pulse shape, these effects manifest as sharp decreases in the magnitude where the velocity of trajectory is greatest (near the equator). These dips partially recover as the velocity of the path slows and the trapping rate into the recaptured dark state increases near the $|+1_g\rangle$ pole. Additionally, loss out of the subspace accumulates over the trajectory causing a gradual decrease in dark state magnitude. A four-state master equation capturing these effects is presented in the Methods and Supplemental Section 1 and reproduces the trajectory in Fig. 2b. Finally, we tomographically reconstruct other trajectories for paths enclosing Φ = 0° to 330° in 30° increments, all revealing similar features. Plotted in Fig. 2c are the inbound trajectories of these paths demonstrating the ability to enclose any wedge angle Φ.

To further qualify the adiabaticity of the path, we measure the photoluminescence (PL) during the STIRAP transition. Non-adiabatic evolution permits excitation to $|A_2\rangle$ and emitted photons, while adiabatic evolution remains dark. The time-resolved PL during STIRAP (Fig. 2d, blue) indicates that the interaction is dark when compared to a non-adiabatic interaction that optically pumps the spin from $|-1_g\rangle$ to $|+1_g\rangle$ and then pumps the spin back to $|-1_g\rangle$ midway through the interaction (Fig. 2d, red). This optical pumping is a form of coherent population trapping (CPT)[30,34]. In fact, the average number of photons emitted during the STIRAP interaction is 9 times fewer than the number of photons emitted at the beginning of the CPT interaction when the population is maximally inverted, indicating that STIRAP is significantly more adiabatic than CPT. Notably, we observe that additional photon emission (Fig. 2d) coincides with a reduction in the dark state vector (Fig. 2b), both consequences of the loss of adiabatic following.

**Optical Accumulation of Berry Phase**

With the ability to enclose loops of arbitrary wedge angle, Φ, on the $|-1_g\rangle$ / $|+1_g\rangle$ Bloch sphere, we extend this STIRAP technique to observe the Berry phase, $\gamma_B$, accumulated on $|-1_g\rangle$ after a cycle has completed. To measure this phase, we exploit the triplet nature of the NV center ground state by using the third state $|0_g\rangle$ as a phase reference. We begin by placing the spin into a fixed $|0_g\rangle$ / $|-1_g\rangle$ superposition. We then use STIRAP to enclose a given Φ on the $|-1_g\rangle$ / $|+1_g\rangle$ subspace with $\tau$ = 1200 ns and $\Omega_R$ = 31 MHz. Phase then accumulates on $|-1_g\rangle$ relative to $|0_g\rangle$, which is measured by performing state tomography (Fig. 3a). This final state has an accumulated phase that is the sum of a dynamic phase, $\eta$, that is constant for all wedge angles and a Berry phase, $\gamma_B$, that scales with the wedge angle as (see Methods):

$$\gamma_B = -\Phi. \tag{1}$$

In Fig. 3b, top, we show the X and Y tomographic projections of the final spin state for positive wedge angles (positive loops) and in Fig. 3b, bottom, we repeat the same trajectories in reverse to enclose negative wedge angles (negative loops). Fitting all these projections to a global model consisting of a fixed dynamic phase, we determine that the acquired phase indeed matches the expected relation of the Berry phase to the wedge angle (Eq. 1). The amplitude of the oscillations in the X and Y projections, defined as the visibility, acts as a measure of the percentage of loops where adiabaticity is preserved. This visibility is reduced from unity as any non-adiabatic transition during the interaction nullifies the intended Berry phase for a given cycle. In Fig. 3, the visibility is limited to ~22%, which is lower than the final dark state magnitude of 65% (in Fig. 2) as non-adiabatic transitions that nullify the Berry phase can nevertheless repopulate the dark state through absorption and decay. Unlike previous microwave demonstrations of Berry phase[4,10,14], the visibility of our Berry phase does not depend on the particular phase enclosed.

To confirm that the origin of the phase is purely geometric, we verify that the total acquired phase is additive when multiple loops are completed. For instance, when positive or negative loops are repeated ($C_{++}$ / $C_{--}$), the net phase accumulation is proportional to the number of loops (Fig. 3c and SI). Similarly, when we perform a positive loop followed directly by a negative loop ($C_{+-}$), the geometric phase is completely cancelled, leaving only the fixed dynamic phase. The visibility decreases for multiple loop repetitions due to additional absorption over the increased interaction time (Supplemental Section 2.3.2).

In addition to the Berry phase, the total phase also consists of a fixed dynamic contribution, $\eta$, that is sensitive to the traversal time and energetics of the interaction. During STIRAP, $\eta$ results from an optical Stark effect[29] that shifts the energy of the dark state relative to the state $|0_g\rangle$, altering the spin's precession in the experimental rotating frame. At two-photon resonance where $\delta = 0$, the dark state does not couple to the light fields and experiences no optical Stark shift; however, it is difficult to precisely set $\delta = 0$ *a priori*, and thus $\eta$ arises from this imprecision. To determine the effect of $\delta$, we measure $\eta$ and extract the optical Stark frequency shift, $\Sigma_\eta(\delta) = \frac{1}{360°}\frac{d\eta(\tau,\delta)}{d\tau}$, isolating where $\Sigma_\eta = 0$ for $\delta = 0$ (Supplemental Section 2.3.3). In Fig. 4a and its inset, we find that $\Sigma_\eta$ scales linearly in $\delta$, with a slope independent of the optical Rabi frequency, $\Omega_R$. Perturbation theory in small $\delta$ reveals that $\Sigma_\eta$ depends only on the ratio of the field amplitudes $\Omega_{-1}(t)/\Omega_{+1}(t)$, and yields an expected relation of $\Sigma_\eta = 0.55\,\delta$ given our pulse shape (Fig. 4a, inset, dashed line), which matches well to the experimental result (Supplemental Section 2.3.3). Unlike the dynamic phase, which requires fine control of $\delta$ and $\tau$, the Berry phase has no dependence on either of these parameters as long as STIRAP remains in the adiabatic regime.

**Limits and Robustness of Berry Phase**

To isolate the effects of adiabaticity and control noise on the Berry phase, we implement a Hahn echo sequence[10,14] to cancel the dynamic phase. We begin with a positive loop that encloses $\gamma_{B,1} = -\Phi$ and accumulates a total phase $\xi_1 = \gamma_{B,1} + \eta$. A microwave π-pulse then flips the sign of the previously accumulated phase, after which we perform a negative loop that encloses $\gamma_{B,2} = \Phi$ and accumulates the same dynamic phase $\eta$, leading to an additional accumulation of $\xi_2 = \gamma_{B,2} + \eta$. This results in a total phase accumulation of $\xi = -\xi_1 + \xi_2 = 2\Phi$ with no contribution from the dynamic phase.

To understand where this geometric control breaks down, we examine the visibility of the echoed Berry phase as a function of the traversal time, $\tau$, and the Rabi frequency, $\Omega_R$ (Fig. 4b). For a given $\Omega_R$, we find a sharp decrease in the visibility where adiabaticity is completely lost for short $\tau$. Likewise, gradual reduction in the visibility for longer $\tau$ is due to decoherence that increases the probability of cycling through $|A_2\rangle$ for a given iteration, thus obfuscating the Berry phase. As we increase $\Omega_R$ to 64 MHz, we achieve visibilities as high as 51% and adiabatic interaction times as short as $\tau \sim$ 250 ns. Faster adiabatic evolution is enabled by increasing $\Omega_R$ as the energy gap between dark and non-dark states expands. Modelling of these general trends using our four-state master equation approach is presented in Supplemental Section 1.4. This geometric control with STIRAP represents a 100-fold speedup over the previous atomic demonstration[17].

Furthermore, as the Berry phase arises from global geometric properties of the state evolution, it offers a degree of robustness to noises that act locally on the trajectory. To investigate, we introduce simulated noise onto the input parameters controlling the polar $\theta(t)$ and azimuthal $\phi(t)$ angles for our loops. The two types of noise, $\delta\phi(t)$, acting perpendicular to the ideal path, and $\delta\theta(t)$, acting parallel to the ideal path, physically correspond to fluctuations in the relative phase and amplitude, respectively, of the two laser fields controlling STIRAP (Fig. 5a). We measure the standard deviation $\sigma_{\gamma B}$ of the distribution of Berry phases realized from 250 unique instances of noisy paths[4]. The noises conform to an Ornstein-Uhlenbeck process with a Lorentzian frequency bandwidth $\Delta\nu$ = 3 MHz and a Gaussian distribution of amplitudes with standard deviation $s_i$ ($i = \theta, \phi$). In Fig. 5b, we plot the distributions (including broadening by photon collection statistics) arising from a noise amplitude of $s_\phi$ = 8° for ideal loops enclosing four disparate Berry phase angles. We find these distributions remain constant regardless of the intended Berry phase. This feature is conducive to practical protocols, as the sensitivity to noise fluctuations does not depend on the given $\Phi$ (Supplemental Section 2.4.4), unlike other approaches[4,10] where larger Berry phases are more susceptible to noise. This arises from the path-length preserving nature of our trajectories that are conveniently accessed by STIRAP.

In Fig. 5c, we examine the impact of increasing the amplitude of the two different types of noise. Consistent with the expectation that parallel noise does not change the enclosed solid angle, the Berry phase remains minimally dephased for increased $\delta\theta$ noise. However, larger noise amplitudes in $\delta\theta$ reduce the visibility as fewer adiabatic loops are preserved due to non-adiabatic changes introduced by the noise. In the case of perpendicular noise, $\delta\phi$, which modifies the enclosed solid angle, we see an enhanced effect on the distribution of Berry phases. Assuming the dark state adiabatically follows the noisy path, we derive an analytic relationship[3] between the variance in the Berry phase and the noise amplitude $s_\phi$ for our specific trajectory (Supplemental Section 2.4.4),

$$\sigma_{\gamma B}^2 = \frac{s_\phi^2}{2}\left[\frac{1 - e^{-2\pi \Delta\nu \tau}}{(1 + (\Delta\nu \tau)^2)^2} + \frac{\pi \Delta\nu \tau}{1 + (\Delta\nu \tau)^2}\right]. \qquad (2)$$

This variance has no dependence on the wedge angle $\Phi$, but only depends on the product $\Delta\nu \tau$, a measure of the number of noise oscillations per cycle. The same derivation predicts insensitivity to $s_\theta$ to first order. The intrinsic $\sigma_{\gamma B}$ can be estimated from the shot-noised broadened standard deviations $\hat{\sigma}_{\gamma B, echo}$ by subtracting the estimated photon collection shot noise contribution in quadrature and dividing by two to account for the two loops traversed in the echo measurement (Supplemental Section 2.4.3). In Fig. 5d, we confirm that $\sigma_{\gamma B}$ is strongly robust to $\delta\theta$ noise, while its dependence on $\delta\phi$ noise matches well to the expected result $\sigma_{\gamma B} = 0.64\, s_\phi$ from Eq. 2 (solid line) using the experimental parameters $\Delta\nu$ = 3 MHz and $\tau$ = 1200 ns. In contrast to dynamic phase, adiabatic geometric phase becomes increasingly robust to noise as the traversal time increases, as can be seen in $\sigma_{\gamma B}^2 \to \frac{\pi\, s_\phi^2}{2\, \Delta\nu\, \tau}$ in the limit of $\Delta\nu\, \tau \gg 1$. In Fig. 5e, we display the estimated $\sigma_{\gamma B}$ as a function the STIRAP traversal time $\tau$ for measurements at $\Delta\nu$ = 3 MHz and constant noise amplitude $s_\phi$ = 14°. These measurements clearly demonstrate the predicted $\sigma_{\gamma B} \sim \tau^{-1/2}$ scaling that is the hallmark of noise resiliency for geometric phases.

**Conclusions and Discussion**

We demonstrate an all-optical approach to accumulate Berry phase in a solid-state system that enables independent, geometric manipulatation of individual qubits with diffraction-limited spatial resolution. Using the $|A_2\rangle$ $\Lambda$ system of the NV center in diamond, we control the adiabatic passage of a dark state, understand the mechanisms that limit the successful enclosure of Berry phase, and characterize the nature of its robustness to noise. Due to imperfect initialization and loss mechanisms, the experimental Berry phase visibilities peak at 51%, corresponding to an estimated peak state fidelity of 73%; this fidelity could be improved in a more isolated $\Lambda$ system allowing for higher optical driving powers (Supplemental Sections 1.4 and 1.5). Extensions to this technique could be realized by harnessing other solid-state $\Lambda$ systems, such as in the silicon-vacancy (SiV) in diamond[35,36] with its strong zero-phonon line emission, important for photonic applications[37]. Alternatively, adding another optical field to actively control the third ground state level (e.g. the reference level $|0_g\rangle$) in a solid-state tripod system provides an avenue for an all-optical set of universal geometric single qubit gates[6,17–19]. The prevalence of $\Lambda$ and tripod energy structures make these techniques extendable to a variety of solid-state qubits, including color centers[35,36,38–40], transition metal[41] or rare-earth ions[22,42], and quantum dots[43], existing in materials[44] promising for a broad range of photonic technologies.

**Methods**

*Experimental Setup*

The experiments in this work use an electronic grade diamond substrate purchased from Element Six, measuring 2 x 2 x 0.5 mm. All NV centers present in the sample were naturally formed during the growth process. We lithographically patterned Ti:Au (10 nm Ti, 100 nm Au) short-terminated waveguides on the surface to provide on-chip microwave control of the NV centers. The sample is thermally sunk inside a liquid helium flow cryostat held at 8 K. The short-terminated waveguide is wirebonded to a microwave line within the cryostat and connected to the signal generators via a coaxial port. The cryostat serves as the sample chamber for a confocal microscopy setup designed to study individual NV centers. The NV center studied had a natural optical linewidth of ~100 MHz with an orbital strain splitting around 7.4 GHz. An applied external magnetic field of 117 G splits $|-1_g\rangle$ and $|+1_g\rangle$ by 655 MHz, and the combination of the natural strain and applied magnetic field split the $|A_2\rangle$ and $|A_1\rangle$ excited states by ~2.9 GHz.

The confocal microscopy setup consists of a 532 nm laser to re-ionize the NV$^-$ charge state and initialize to the $m_S$ = 0 spin state, a tunable 637 nm laser tuned to the $|E_Y\rangle$ transition for readout of the spin state[45], and second tunable 637 nm laser fiber coupled to an electro-optic modulator tuned to the $|A_2\rangle$ transition for the STIRAP interaction. The electro-optic modulator is driven by a signal generator tuned to 655 MHz, the splitting of $|-1_g\rangle$ and $|+1_g\rangle$, creating sidebands on the laser to drive the Λ transitions. All lasers are controlled using acousto-optic modulators for nanosecond timescale pulsing. All three lasers pass through individual polarization optics, and are combined using beamsplitters and dichroic mirrors. The combined beam is eventually focused onto the sample using a 0.85 NA 100x objective that is aberration-corrected for the cryostat window. The red-shifted phonon sideband of the NV center's PL is spectrally filtered through a series of dichroic mirrors and bandpass filters and then counted in a silicon avalanche photodiode. Those counts are binned via a series of logic switches, and then summed by either a time-correlated counting card (Fig. 2d) or a data acquisition card (rest of Fig. 2, Figs. 3-5).

In addition to the 655 MHz applied to the EOM, additional microwave frequencies are needed for characterization. For the Berry phase measurements (Figs. 3-5), a second signal generator provides on-chip microwaves tuned to 2.550 GHz, the splitting of the ground state $|0_g\rangle$ and $|-1_g\rangle$ levels. However, for the STIRAP path evaluation measurements (Fig. 2), three colors of microwaves are required. In this case, the second signal generator provides microwaves tuned to 3.205 GHz, the splitting of $|0_g\rangle$ and $|+1_g\rangle$, while a frequency mixer combines the two initial frequencies to provide the third frequency, 2.550 GHz, the $|0_g\rangle$ / $|-1_g\rangle$ splitting. To phase-control the microwaves, we use the internal IQ modulation functionality of both signal generators. All timing and pulse sequences (Supplemental Sections 2.2.2, 2.3.1, and 2.4.1) are controlled with a 1 GS/s arbitrary waveform generator.

*Theoretical methods*

The rotating frame Hamiltonian describing the Λ system and optical fields within the NV center level structure is:

$$H = \frac{h}{2}\begin{pmatrix} 0 & 0 & 0 & 0 \\ 0 & 0 & 0 & \Omega_{-1}(t) \\ 0 & 0 & 2\delta & \Omega_{+1}(t)e^{i\phi(t)} \\ 0 & \Omega_{-1}(t) & \Omega_{+1}(t)e^{-i\phi(t)} & 2\Delta \end{pmatrix} \quad (3)$$

where the matrix representation is given in the basis $\{|0_g\rangle, |-1_g\rangle, |+1_g\rangle, |A_2\rangle\}$. The master equation in Lindblad form is given by:

$$\dot{\rho} = -i[H,\rho] + \sum_k \left(L_k \rho L_k^\dagger - \frac{1}{2}L_k^\dagger L_k \rho - \frac{1}{2}\rho L_k^\dagger L_k\right) \quad (4)$$

where $L_k$ denote the Lindblad operators describing dissipative processes. These include experimentally estimated relaxation times from $|A_2\rangle$ to $|-1_g\rangle$ of ~31 ns, $|A_2\rangle$ to $|+1_g\rangle$ of ~24 ns, $|A_2\rangle$ to the reference state $|0_g\rangle$ of ~104 ns, an orbital dephasing rate of 7 ns[31], and a fitted phenomenological spin dephasing rate of 2.25 μs. The optical fields in the simulation are described by $\Omega_R = 31\ MHz$ and $\Delta = 60\ MHz$. See Supplemental Section 1 and 2.1.2 for more details.

*Berry Phase*

The dark state in our system is:

$$|D\rangle = \cos\left(\frac{\theta(t)}{2}\right)|-1_g\rangle - \sin\left(\frac{\theta(t)}{2}\right)e^{i\phi(t)}|+1_g\rangle \quad (5)$$

where $\theta(t) = 2\tan^{-1}\frac{\Omega_{-1}(t)}{\Omega_{+1}(t)}$ and $\phi(t)$ is the phase relation between the driving fields. The total Berry phase accumulation over an adiabatic trajectory of a dark state is given by[19],

$$\gamma_B = i\int_{R_i}^{R_f}\langle D|\nabla_{\bar{R}}|D\rangle \cdot d\bar{R}, \quad (6)$$

where the vector $\bar{R} = \begin{pmatrix}\theta\\\phi\end{pmatrix}$ describes the surface of the Bloch sphere. Substituting our dark state, the Berry phase simplifies to:

$$\gamma_B = -\oint \sin^2\left(\frac{\theta(t)}{2}\right)d\phi \quad (7)$$

where the integral is taken over the closed loop on the Bloch sphere. To determine the accumulation from our specific path, non-zero contributions to the integral only occur where the phase shift occurs, $\Delta\phi = \Phi$ at the $|+1_g\rangle$ pole, $\theta$ = 180°, and $\Delta\phi = -\Phi$ at the $|-1_g\rangle$ pole, $\theta$ = 0°. From this, we determine a Berry phase of:

$$\gamma_B = -\Phi \tag{8}$$

**Acknowledgements**

We thank C. P. Anderson, B. B. Buckley, D. J. Christle, and C. F. de las Casas for thoughtful discussions and H. L. Bretscher for experimental assistance. This work was supported by the Air Force Office of Scientific Research, the National Science Foundation, and the German Research Foundation (SFB 767).

**Author Contributions**

C.G.Y., F.J.H, and B.B.Z. performed the experiments. A.A. and G.B. developed the theoretical modeling. All authors contributed to the data analysis and writing of the paper.

**Figures**

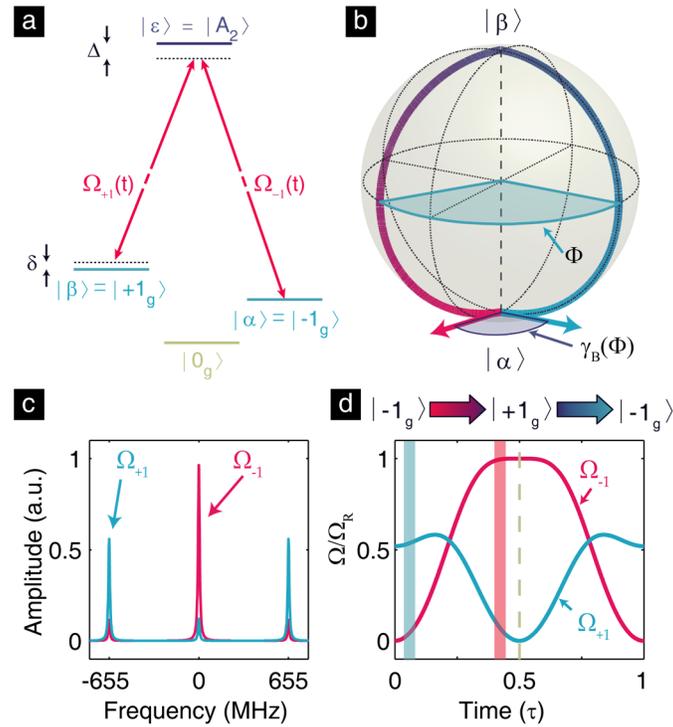

**Figure 1 | Driving the $|A_2\rangle$ Λ System**

**a)** Λ system within the NV center level structure consisting of $|-1_g\rangle$ and $|+1_g\rangle$ coupled to the spin-orbit excited state $|A_2\rangle$ by two optical driving fields $\Omega_{+1}(t)$ and $\Omega_{-1}(t)$, with one-photon detuning, $\Delta$, and two-photon detuning, $\delta$. **b)** Time trace (red->blue gradient) of state transfer through STIRAP on the Bloch sphere. The gradient red trajectory indicates transfer from a dark state $|\alpha\rangle$ to a dark state $|\beta\rangle$. Returning the dark state to $|\alpha\rangle$ along a different longitude (gradient blue) encloses a wedge angle, $\Phi$. Berry phase, $\gamma_B$, accumulates on $|\alpha\rangle$, proportional to $\Phi$. **c)** Frequency harmonics of the electro-optic modulator split by 655 MHz, to drive both transitions in the Λ system. Assignment of harmonics to transitions is indicated. **d)** Example trace of the relative optical Rabi frequency of the driving fields, $\Omega_{+1}(t)$ (blue) and $\Omega_{-1}(t)$ (red), as a function of time showing the movement of the dark state from the $|-1_g\rangle$ pole ($t = 0$) to the $|+1_g\rangle$ pole ($t = \frac{\tau}{2}$) and back ($t = \tau$) during a STIRAP interaction. The pump amplitudes in c) correspond to the times highlighted by the red and blue linecuts in d).

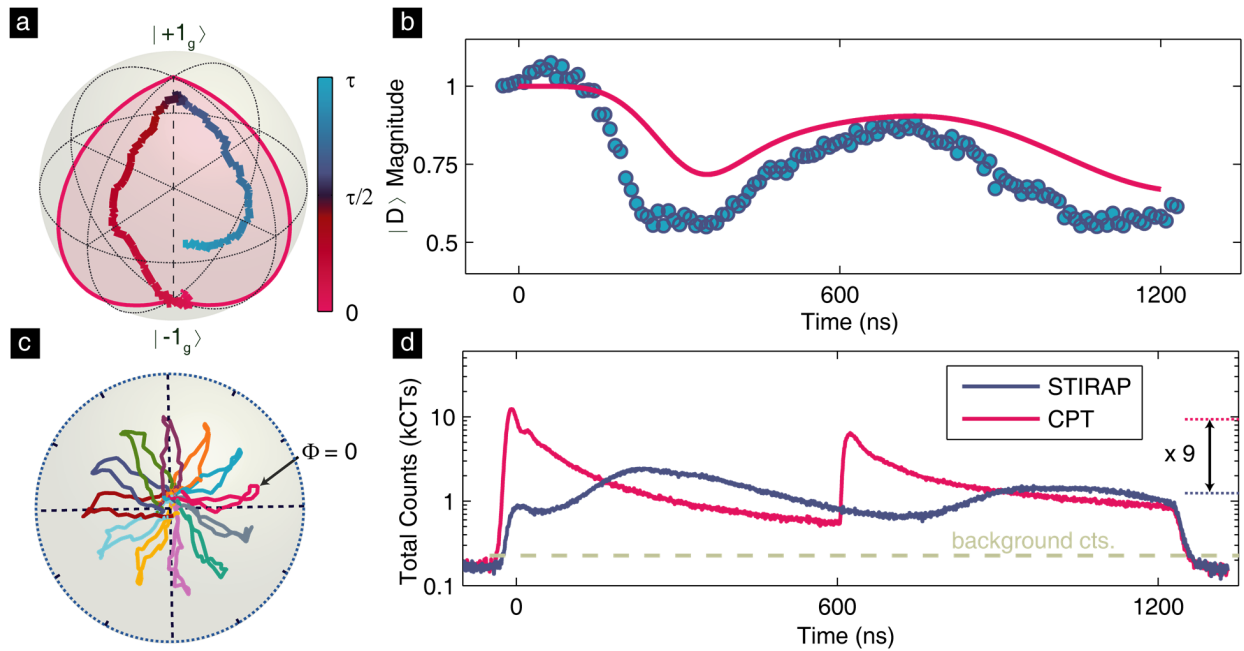

**Figure 2 | Characterizing Phase-Controlled STIRAP in the NV Center**

**a)** Tomographically reconstructed path of the spin on the $|-1_g\rangle$ / $|+1_g\rangle$ Bloch sphere during the STIRAP interaction moving from $|-1_g\rangle$ to $|+1_g\rangle$ and back enclosing Φ = 120° ($\tau$ = 1200 ns). The shaded region shows the solid angle enclosed by the ideal trajectory. **b)** The magnitude of the dark state during the same STIRAP interaction shows a sharp decrease around the equator due to occupation of non-dark states from non-adiabatic transitions. The gradual decrease in the magnitude over time is attributed to loss out of the $|-1_g\rangle$ / $|+1_g\rangle$ subspace into $|0_g\rangle$. The red curve is calculated from the master equation simulation and is plotted for comparison. **c)** Inbound trajectories of STIRAP loops enclosing Φ = 0° to 330° in 30° steps indicating full Bloch sphere control over cyclic paths. **d)** Time-resolved photoluminescence (PL) during an experimentally identical STIRAP interaction as in a) and b), plotted against a non-adiabatic CPT interaction on a log scale. The relative darkness of the PL during STIRAP indicates that it is a largely adiabatic evolution. Comparing b) and d), we see the PL during STIRAP peaks when the state is near the equator during both outbound and inbound trajectories, corresponding to excitation as a result of non-adiabatic following of the path velocity.

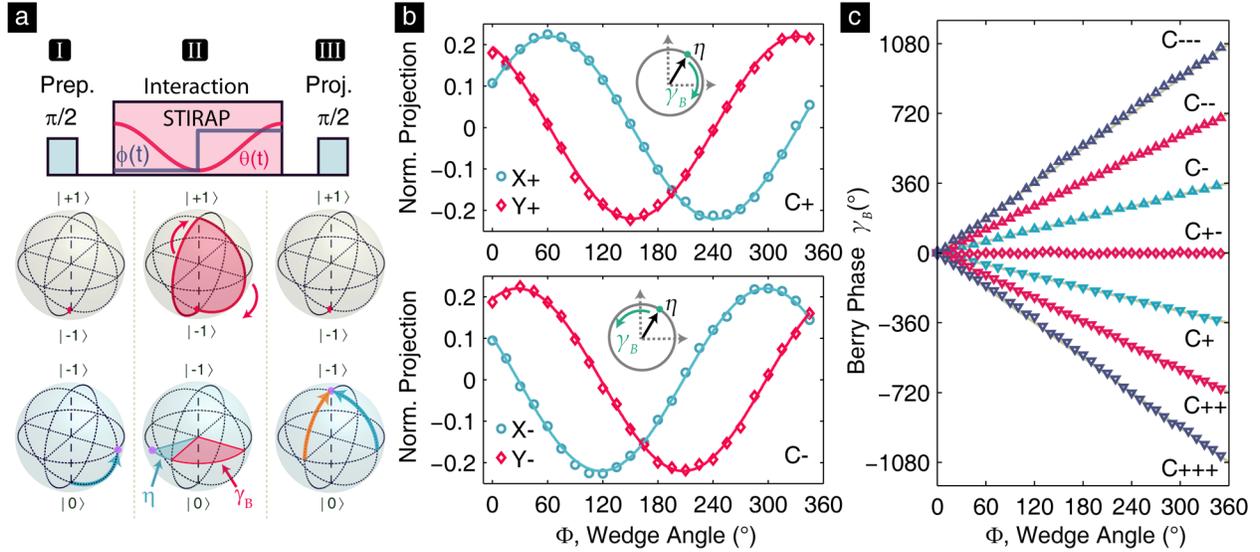

**Figure 3 | Optically Accumulated Berry Phase**

**a)** Pulse sequence to measure Berry phase accumulated during STIRAP interaction (red and indigo curves denote the $\theta(t)$ and $\phi(t)$ trajectories). I. Prepare spin in fixed $|0_g\rangle \,/\, |-1_g\rangle$ superposition using microwave techniques. II. Loop the spin on the $|-1_g\rangle \,/\, |+1_g\rangle$ Bloch sphere enclosing $\Phi$. Phase will accumulate on $|0_g\rangle \,/\, |-1_g\rangle$ Bloch sphere corresponding to a combination of a fixed dynamic phase, $\eta$, and a varying Berry phase, $\gamma_B$, which is a function of $\Phi$ (Eq. 1). III. A final projection pulse reads out the accumulated phase on $|0_g\rangle \,/\, |-1_g\rangle$ Bloch sphere through state tomography. **b)** Resulting X and Y projections of the accumulated phase for a positive and negative loop. Projections fit to $X_\pm = A\cos(\eta \pm \gamma_B(\Phi))$ and $Y_\pm = A\sin(\eta \pm \gamma_B(\Phi))$ where A is the visibility and $\gamma_B(\Phi) = -\Phi$. **c)** Measured Berry phase when multiple loops of wedge angle $\Phi$ are traversed. Single loops (blue), double loops (red), and triple loops (indigo) indicate the expected additive behavior. A loop consisting of a positive loop followed by a negative loop of $\Phi$ (C+-) indicates full cancellation of the Berry phase. The solid lines are guides to the eye ($\gamma_B = N\Phi$ for N = -3,-2,...,3). Errors in b) and c) are approximately the size of the points.

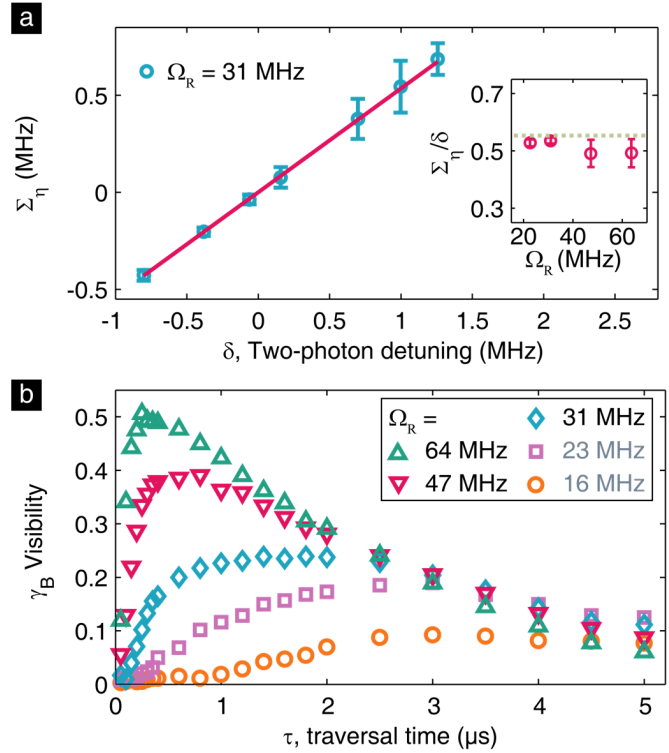

**Figure 4 | Exploring the Dynamic and Berry Phases**

**a)** The dark state energy shift, $\Sigma_\eta$, as a function of the two photon-detuning, $\delta$, for a STIRAP interaction of maximum optical Rabi frequency, $\Omega_R$ = 31 MHz. This energy shift multiplied by the traversal time $\tau$ determines the total dynamic phase accumulation. The inset displays the ratio of $\Sigma_\eta/\delta$ as a function of $\Omega_R$. The dashed line indicates the expected behavior of $\frac{\Sigma_\eta}{\delta} = 0.55$ for small two-photon detuning. Errors represent 95% confidence intervals. **b)** The visibility of the Berry phase, $\gamma_B$, as a function of traversal time, $\tau$, for different $\Omega_R$. A sharp turn-on for small $\tau$ indicates the adiabatic limit, while the gradual decrease in visibility for longer $\tau$ is due to accumulated excitation to $|A_2\rangle$. Errors are smaller than the point size.

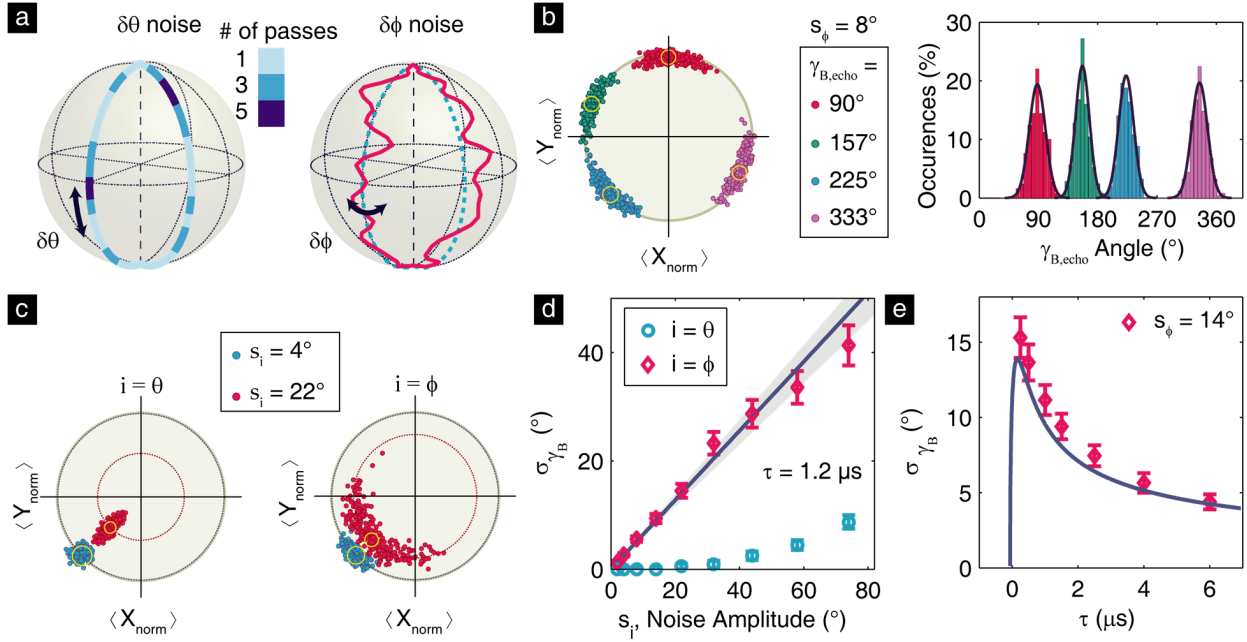

**Figure 5 | Noise Robustness of Berry Phase**

**a)** Illustration of $\delta\theta$ (left) and $\delta\phi$ (right) noise on Bloch sphere. $\delta\theta$ noise is parallel to the path and does not affect the enclosed solid angle, while $\delta\phi$ noise is perpendicular to the path and affects the enclosed solid angle. **b)** Angular distributions (broadened by photon collection statistics) of measured Berry phases for a specific noise amplitude, $s_\phi = 8°$, with intended phases: $\gamma_{B,echo}$ = 90° (red), 157° (green), 225° (blue), and 333° (purple), plotted on the equatorial slice of Bloch sphere (left) and binned into histograms with bin size of 6° (right). Projections are normalized by zero-noise case, for b) and c). Yellow circle indicates the 95% confidence interval of photon collection shot noise about the intended $\gamma_B$. The width of the distributions is independent of the intended Berry phase. **c)** Shot-noise broadened angular distributions of $\gamma_B$ = 225° for both $\delta\theta$ and $\delta\phi$ noise at noise amplitudes $s_i = 4°$ (blue) and 22° (red). Dashed circle indicates mean visibility, $\langle\sqrt{X^2+Y^2}\rangle$, of the distribution for $s_i = 4°$ (blue) and 22° (red). The smaller magnitude of the visibility indicates fewer adiabatic loops are preserved. **d)** Estimated standard deviation, $\sigma_{\gamma B}$, of intrinsic distributions vs. noise amplitude $s_i$ for both $\delta\theta$ (blue) and $\delta\phi$ (red). Grey shaded region is 95% confidence interval on the experimental slope. **e)** Estimated $\sigma_{\gamma B}$ showing a $\tau^{-1/2}$ decrease as the STIRAP traversal time, $\tau$, increases for a constant noise amplitude $s_\phi = 14°$. Errors in d) and e) represent 95% confidence intervals. Indigo lines in d) and e) are the predicted behavior for $\sigma_{\gamma B}$ using experimental parameters and Eq. 2 of the main text.